\documentclass{article}
\usepackage{epsfig}
\usepackage{multicol}

\def\title#1{\begin{center}\begin{minipage}{\textwidth}
\center\uppercase{#1}\endcenter\end{minipage}\end{center}}

\def\author#1{\vspace*{-1.5ex}%
{\topsep 3pt\center\small\scshape#1\endcenter}\vspace*{-1pt}}

\def\affil#1{\vspace*{-0.8ex}{\topsep 0pt\center\footnotesize%
\def\baselinestretch{1.0}%
#1\endcenter}}

\def\and{\vspace{0pt}{\topsep 0pt\center {\sc and}\endcenter}\vspace{3pt}}

\def\fig1{
\epsfig{width=3.5in,file=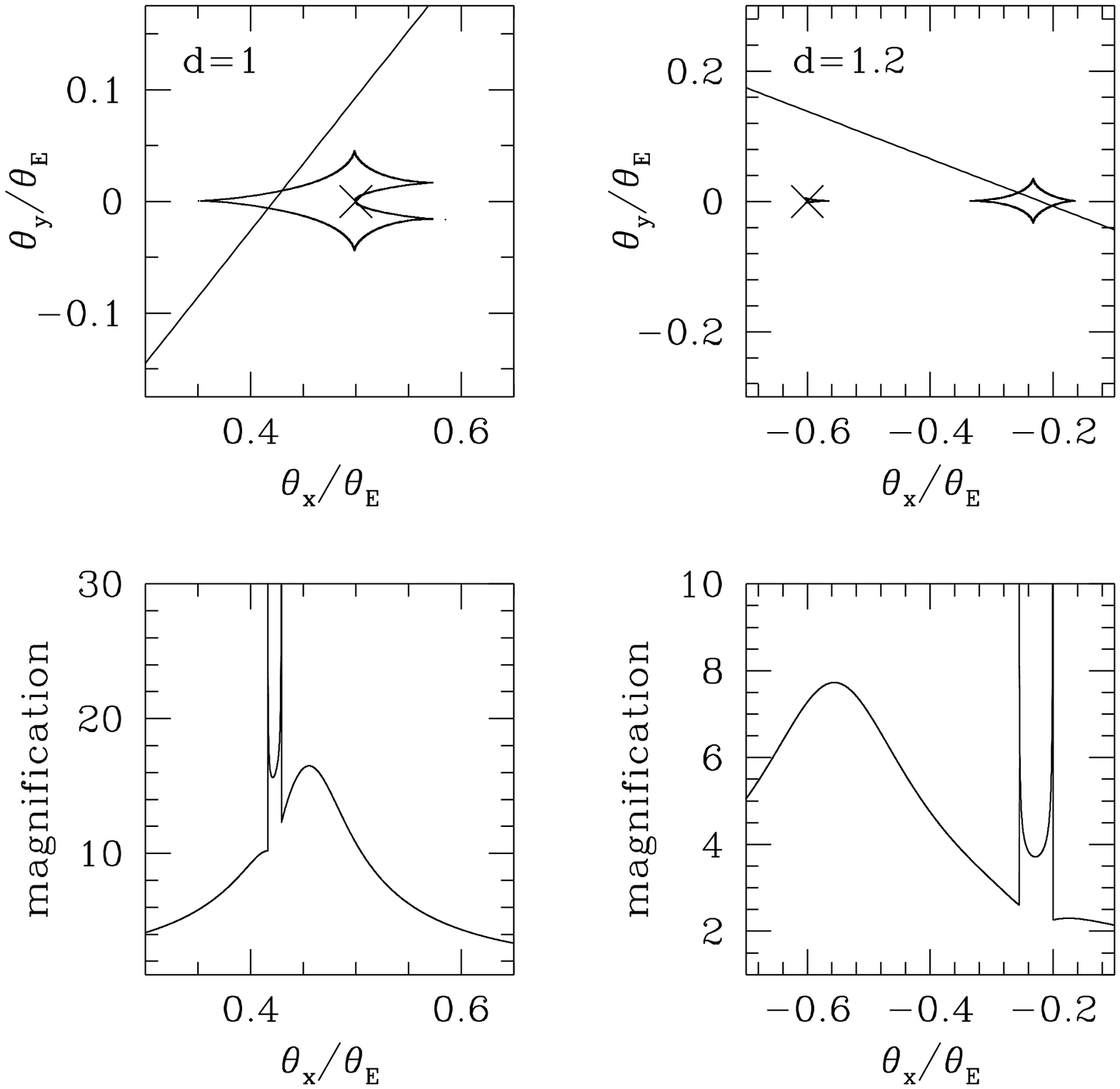} 
\vbox{\small \baselineskip2pt
  Figure 1: Microlensing events due to a planetary system.  We illustrate the
  trajectory of a source star over the caustic curves of a lens star and its
  planet.  We then illustrate the observed magnification of the source star
  along its trajectory.  The cross indicates the star's position, whereas the
  planet lies off the plots at $-0.5$ and $0.6$, respectively.  The
  plots are in units of the Einstein angle $\theta_E$, which is the
  characteristic angular scale of a microlensing event.  Also shown is the
  star--planet separation $d$ in units of $\theta_E$.\\~}
}

\begin{document}
\title{Searching for extragalactic planets}
\author{Edward A. Baltz}
\affil{Department of Physics, University of 
California, Berkeley, CA 94720-3411, USA
\\{\rm Email: \tt eabaltz@astron.berkeley.edu}}
\and
\author{Paolo Gondolo}
\affil{Max-Planck-Institut f\"{u}r Physik, F\"{o}hringer Ring
6, D-80805 M\"{u}nchen, Germany
\\{\rm Email: \tt gondolo@mppmu.mpg.de}}

\begin{abstract}
  Are there other planetary systems in our Universe? Indirect evidence has
  been found for planets orbiting other stars in our galaxy: the gravity of
  orbiting planets makes the star wobble, and the resulting periodic Doppler
  shifts have been detected for about a dozen stars~\cite{1}. But are there
  planets in other galaxies, millions of light years away?  Here we suggest a
  method to search for extragalactic planetary systems: gravitational
  microlensing of unresolved stars.  This technique may allow us to discover
  planets in elliptical galaxies, known to be far older than our own Milky Way,
  with broad implications for life in the Universe. 
\end{abstract}

\begin{multicols}{2}

The phenomenon of gravitational microlensing is as follows.  A massive object,
be it a black hole, planet, star, etc., passes very near to the line of sight
to a star being monitored.  The gravity of the massive object bends the
starlight (gravitational lensing), producing multiple images of, and
magnifying, the monitored star for a short time. The multiple images can not be
resolved (hence the prefix micro), but the amplification of the light intensity
can be detected.  The amplification has a well-defined characteristic temporal
behaviour, allowing such events to be distinguished from other possible
intensity changes such as those due to variable stars~\cite{2-4}.

In the case where the lens itself is a binary object, the microlensing
lightcurve can be strongly affected, exhibiting short periods of very large
magnification, coming in pairs~\cite{5-9,10}. Gravitational lensing events
of this type involving binary stars have been observed by the MACHO and EROS
teams in programmes monitoring several millions of stars~\cite{11-15}.
The short duration large magnification events are referred to as caustic
crossings.  The magnification along these caustic curves is formally infinite,
and in practice quite large, in excess of ten.  Caustics are well known in
optics, and can be seen, for example, as the oscillating patterns of bright
lines on the bottom of a swimming pool.

The utility of such events in searching for planetary systems is clear: a solar
system can be described to first approximation as a binary object, as is the
case of our own solar system, consisting primarily (in mass) of the Sun and the
planet Jupiter.  In Figure 1, we show two example lightcurves of microlensing
events, together with the trajectories of the source stars relative to the
caustic curves.  In both cases the star has a companion
one one-thousandth as massive, like the Sun--Jupiter system.

Previous work has shown that planets might be detected in microlensing events
in the bulge of the Milky Way galaxy, or in the Small and Large Magellanic
Clouds, which are small galaxies in orbit around the Milky Way~\cite{10}.
Stars in the bulge and in the Magellanic clouds can be resolved easily, and
surveys routinely monitor of order ten million stars for microlensing events.
Evidence for a planet orbiting a binary star system in the Milky Way bulge has
recently been presented in a joint publication of the MPS and GMAN
collaborations~\cite{16}.

In order to observe planetary systems in more distant galaxies, we must resort
to a technique known as pixel microlensing~\cite{17-18}. Individual stars
in distant galaxies can not be resolved, but this does not invalidate the
method.  Each pixel of a telescope camera collects light from a number of stars
in the distant galaxy.  If a single star is magnified due to gravitational
microlensing, the pixel will collect more light.  Of course, the light from
other stars on the pixel makes a magnification more difficult to observe, but
the technique works in practice~\cite{19-26}.

The pixel microlensing surveys that have been undertaken observe the Andromeda
Galaxy (M31), the nearest large galaxy to the Milky Way.  These surveys have
used ground based telescopes.  The bulge of M31 is quite dense, which allows a
high probability of microlensing events. We have calculated the rate of
planetary events observing M31 with a telescope like the
Canada-France-Hawaii telescope (CFHT) on Mauna Kea~\cite{27}. We assume
that every star has a companion that is one one-thousandth as massive, just
like the Sun and Jupiter.  Furthermore, we assume, as is true for known binary
stars, that the distribution of orbital periods is such that ten percent of
such systems lie in each decade of period, from a third of a day to ten million
years.  This gives a 10\% probability that a star has a companion between one
and five AU (Astronomical Unit, the distance between the Earth and the Sun), in
rough agreement with the observational findings of Marcy et al.~\cite{1}.
With a long--term monitoring program observing every night for the five months
that M31 is visible in Hawaii over a period of eight years, we expect to
observe about one planetary event.

To increase the chances to detect planetary systems in distant galaxies, we
require a space telescope such as the proposed Next Generation Space Telescope,
to be launched around 2007.  This will be a large (about 8 metres in diameter)
infrared telescope at a Lagrange point \noindent\fig1\noindent 
of the Earth--Moon system, and it will
be more than ten times as sensitive as the Hubble Space Telescope.

We have calculated the rate of events we might detect with the NGST observing
the giant elliptical galaxy M87 in the Virgo cluster, at a distance of fifty
million light--years.  With the same assumptions as the M31 calculation, we
find that an NGST survey of two month's duration, taking one image each day,
should be able to detect of order three planetary systems.  We find that such a
survey is most sensitive to events where the separation between caustic
crossings is about five days. An alert system for microlensing events would
allow more frequent measurements of the light curve during the caustic
crossings, with the possibility of determining the orbital parameters of the
planetary system.

We now make some comments about the dependence of the observed rate of
microlensing events on the various physical parameters.  As long as the size of
the observed galaxy is small compared to the distance to it, the fundamental
rate of events is constant.  However, a more distant galaxy will appear
smaller, with more stars per pixel, which decreases the rates.  On the other
hand, if the galaxy is too close, many images must be taken in order to monitor
enough stars, requiring more telescope resources.  For the pixel scale of the
NGST, M87 is at a fairly optimal distance.

We have shown that pixel microlensing may be used to detect extragalactic
planetary systems.  This is the most promising available technique for
discovering planets outside our own galaxy.  This is especially interesting in
that we might find planets in elliptical galaxies such as M87, known to contain
considerably less heavy elements than our own spiral galaxy. The discovery of
an extragalactic planet in a galaxy very different from ours would have broad
implications for the origin of life in the Universe.

\end{multicols}

\end{document}